\documentclass{emulateapj}
\defcitealias{My2009ApJ}{Paper~I}

\slugcomment{{\it ApJL} accepted}

\usepackage{ulem}
\usepackage{color}
\usepackage{url}

\shorttitle{Short Lifetime of Protoplanetary Disks}
\shortauthors{Yasui et al.}

\begin{document}

\title{Short Lifetime of Protoplanetary Disks in Low-metallicity
Environments\altaffilmark{1}}
\altaffiltext{1}{Based on data collected at Subaru Telescope, which
is operated by the National Astronomical Observatory of Japan.}

%%% \author{Genevieve J. Graves\altaffilmark{1,2,3} \&
%%%   S. M. Faber\altaffilmark{1}}
%%% 
%%% \altaffiltext{1}{UCO/Lick Observatory, Department of Astronomy and
%%% Astrophysics, University of California, Santa Cruz, CA 95064, USA}
%%% \altaffiltext{2}{Department of Astronomy, University of California,
%%% Berkeley, CA 94720, USA; graves@astro.berkeley.edu}
%%% \altaffiltext{3}{Miller Fellow}

\author{Chikako Yasui\altaffilmark{2, 3}, Naoto
Kobayashi\altaffilmark{3}, Alan T. Tokunaga\altaffilmark{4}, Masao
Saito\altaffilmark{2}, and Chihiro Tokoku\altaffilmark{5}}
% \footnote{Also at: Subaru
%Telescope, National Astronomical Observatory of Japan, 650 North A`ohoku
%Place, Hilo, HI 96720, USA.}}
%
%\altaffiltext{3}{Miller Fellow}
\altaffiltext{2}{National Astronomical Observatory of Japan 2-21-1
Osawa, Mitaka, Tokyo, 181-8588, Japan: \email{ck.yasui@nao.ac.jp}
}
%
%ALMA Project, National Astronomical Observatory of
%Japan, 2-21-1 Osawa, Mitaka, Tokyo 181-8588, Japan}

\altaffiltext{3}{Institute of Astronomy, School of Science, University of Tokyo,
2-21-1 Osawa, Mitaka, Tokyo 181-0015, Japan}
%\affil{Institute of Astronomy, School of Science, University of Tokyo,
%2-21-1 Osawa, Mitaka, Tokyo 181-0015, Japan}
%
%\email{ck$_-$yasui@ioa.s.u-tokyo.ac.jp}

%\author{Alan Tokunaga\altaffilmark{4,5}}
%\author{Alan T. Tokunaga}
\altaffiltext{4}{Institute for Astronomy, University of Hawaii, 2680 Woodlawn
Drive, Honolulu, HI 96822, USA}
%\affil{Institute for Astronomy, University of Hawaii, 2680 Woodlawn
%Drive, Honolulu, HI 96822, USA}

%\and

%\author{Chihiro Tokoku} %\altaffilmark{4,5}}
\altaffiltext{5}{Astronomical Institute, Tohoku University, Aramaki,
Aoba, Sendai 980-8578, Japan}

%\affil{Astronomical Institute, Tohoku University,
%Aramaki, Aoba, Sendai 980-8578, Japan}

%%%%%%%%%%%%%%%%%%%%%%%%%%%%%%%%%%%%%%%%%%%%%%%%%%%%%%%%%%%%%%%%%%%%%%%%%%%%%%%

\begin{abstract}
\noindent
We studied near-infrared disk fractions of six young clusters in the
low-metallicity environments with [O/H$] \sim -0.7$ using deep $JHK$
images with Subaru 8.2\,m telescope. We found that disk fraction of the
low-metallicity clusters declines rapidly in $<$1\,Myr, which is much
faster than the $\sim$5--7\,Myr observed for the solar-metallicity
clusters, suggesting that disk lifetime shortens with decreasing
metallicity possibly with an $\sim$$10^Z$ dependence.
Since the shorter disk lifetime reduces the time available for planet
formation, this could be one of the major reasons for the strong
planet--metallicity correlation.
Although more quantitative observational and theoretical assessments are
 necessary, our results present the first direct observational evidence
 that can contribute to explaining the planet--metallicity correlation.

\end{abstract}

% http://www.journals.uchicago.edu/page/apj/instruct.key.html

\keywords{
Galaxy: abundances -- 
infrared: stars --
open clusters and associations: general --
planetary systems: protoplanetary disks --
stars: formation --
stars: pre-main sequence}

%%%%%  TABLE ALL %%%%%%%%%%%%%%%%%%%%%%%%%%%%%%%%%%%%%%%%%%%%%%%%%%%%%%%%%%%%%

%%%%%%%%%%%%%%%%%%%%%%%%%%%%%%%%%%%%%%%%%%%%%%%%%%%%%%%%%%%%%%%%%%%%%%%%%%%%%%%
%%%%%%%%%%%%%%%%%%  INTRODUCTION  %%%%%%%%%%%%%%%%%%%%%%%%%%%%%%%%%%%%%%%%%%%%%
%%%%%%%%%%%%%%%%%%%%%%%%%%%%%%%%%%%%%%%%%%%%%%%%%%%%%%%%%%%%%%%%%%%%%%%%%%%%%%%

 \section{INTRODUCTION} \label{sec:intro}

Because planets are formed in protoplanetary disks, the lifetime of
protoplanetary disks ($\tau_{\rm disk}$) is thought to be one of the
most fundamental parameters directly connected to the planet formation
probability ($P_{\rm pl}$).
%\citep{Trieloff2006}.
%\citep{Hillenbrand2008}.
%
In particular, in the current standard core-accretion scenario of gas
giant planet formation \citep{{Safronov1969},{Lissauer_PPV}},
%it had been thought that 
a long disk lifetime of more than 10\,Myr was thought to be
necessary to produce the planetesimals that eventually become the rocky
core of the giant planets (e.g., 100\,Myr; \citealt{Lissauer1993}).
However, \citet{Strom1989} studied the time variation of the frequency
of disk-harboring stars in Taurus molecular cloud that have $K$-band
excess and suggested that the disk lifetime is in the range from
$\ll$3\,Myr to $\sim$10\,Myr, which is much shorter than was thought to
be necessary.
After their pioneering work, a notion called the ``disk fraction'',
which is defined as the frequency of near-infrared (NIR) to mid-infrared
(MIR) excess stars (disk-harboring stars) within a young cluster, was
developed and widely used to study the disk lifetime
\citep[e.g.,][]{{Lada1999},{Haisch2001ApJL},{Hernandez2007}}.
%(e.g., Lada 1999, Haisch+ 2001, Hernandez+ 2007).
%
In all studies, the disk fraction is found to decrease as a function of
cluster age, showing that the disk lifetime is about 5--7\,Myr for
nearby young clusters.
Because the observation with NIR/MIR is sensitive to only heated dust
located at the inner disk with a stellocentric distance of $\sim$0.1 to
a few\,AU, there still remained a possibility that the disk lifetime
derived with NIR/MIR excess emission is only for the limited region of
the disk.
However, from the (sub-)millimeter continuum observation of cold dust
($\sim$10\,K) in the outer disk ($\ga$50\,AU) by \citet{Andrews2005},
the lifetime of the outer disk is also found to be similar, thus the
{\it entire disk} is now thought to disappear nearly simultaneously.
After the discovery that $\tau_{\rm disk}$ is less than 10\,Myr, it is
widely accepted that giant planet cores must form in this short time for
any theoretical models \citep{Lissauer_PPV}.

Although as many as over 450 exoplanets are now known
\citep{Schneider2009}, these planets are unexpectedly found to have a
lot of diversity in terms of e.g., mass, orbital period, and
eccentricity.
Perhaps the most telling discovery in exoplanet study is the
``planet--metallicity correlation'', the higher probability of a star
hosting a giant planet with increasing metallicity
\cite[e.g.,][]{Fischer2005},
suggesting that metallicity could be the most crucial parameter for
giant planet formation.
Although this correlation is generally interpreted as a natural
consequence of the core-accretion scenario, the origin of it is still
not well-explained despite detailed analyses by either theoretical
studies assuming solar metallicity or observations of solar metallicity
regions \citep[e.g.,][]{{Ida2004ApJ616},{Wyatt2007}}.
Therefore, it is necessary to directly study protoplanetary disks under
different metallicity conditions to find any clues of the
planet--metallicity correlation and to test theories of it.

Since our Galaxy is known to have a metallicity gradient with lower
metallicity for larger Galactic radius, ranging from $\simeq$$-1$\,dex
to $\simeq$$+0.5$\,dex \citep[e.g.,][]{Rudolph2006}, we can explore
different metallicity regions at large distances with high-sensitivity
observations with large telescopes.
As a first step, we focused on the low-metallicity environment in the
outer region of our Galaxy ($R_g \gtrsim 15$\,kpc) to study a region
with significantly lower metallicity ([O/H$]\sim -1$\,dex).
As a first result, we derived the disk fraction of two clusters in Digel
Cloud 2 at $R_g = 19$\,kpc \citep[][hereafter Paper~I]{My2009ApJ}, which
are low metallicity ([O/H$]\simeq -0.7$) and very young ($\sim$0.5\,Myr
old) clusters.
Disk fractions for both clusters were found to be quite low in spite of
their very young age, suggesting that the disk lifetime in low
metallicity is short.
In this paper, we summarize the disk lifetime in low-metallicity
environments with our all targets in the outer Galaxy and discuss the
implication for the planet--metallicity correlation.

\section{TARGET SELECTION AND OBSERVATION} \label{sec:targets}

We searched the literature for star-forming regions in the outer Galaxy
 and with [O/H$]<-0.5$.
Among the listed 10 candidate star-forming regions, we selected four
regions at $R_g \gtrsim 15$\,kpc and in the second quadrant of the
Galaxy as an initial set of samples, mostly by considering the
visibility from the Northern hemisphere.
We also checked whether the candidate regions have at least one
associated stellar aggregate using Two Micron All Sky Survey (2MASS)
data.
As a result, we selected six clusters (Table~\ref{tab:allRESULTS}): one
cluster in each Sh 2-207 and Sh 2-208 \citep{Bica2003AA}, and two in
each Sh 2-209 \citep{Klein2005} and Digel Cloud 2 \citep{NK2008}.
The average metallicity of all target clusters is $\simeq$$-0.7$\,dex,
which is significantly lower than the solar metallicity.

Deep $JHK_S$-band images of the target star-forming regions were
obtained with the 8.2\,m Subaru telescope equipped with MOIRCS
\citep[Multi-Object InfraRed Camera and Spectrograph;][]{Suzuki2008},
which has a wide field of view ($4' \times 7'$) and uses the MKO NIR
photometric filters \citep{Tokunaga2002}.
The weather condition was photometric with excellent seeing
($\simeq$$0\farcs3$--$0\farcs4$) for the Cloud 2 and Sh 2-209
observations on 2006 September 3 UT, while a little cirrus was present
with seeing of $\sim$1$''$ for the Sh 2-207 and Sh 2-208 observations on
2006 November 8, 2007 November 23, and 2008 January 14 UT.
Total integration time for each wavelength band was $\sim$500--1000\,sec.
After the raw data were reduced with standard IRAF procedures, $JHK$
aperture photometry was performed.
%\footnote{IRAF is distributed by the National Optical Astronomy
%Observatories, which are operated by the Association of Universities for
%Research in Astronomy, Inc., under cooperative agreement with the
%National Science Foundation.}  
%
For most clusters aperture photometry with IRAF apphot was enough even
for such large distances because stellar images are sufficiently
smaller than the stellar separations.
Only for two dense clusters in Sh 2-209, we performed photometry with
point-spread function fitting using IRAF daophot
 with scripts for automated photometry, ``autodao''\footnote{Copyright
(C) 2008-2009 Noriyuki Matsunaga
(\url{http://www.ioa.s.u-tokyo.ac.jp/~nmatsuna/Japanese/software/autodao.html})}.

Photometric standard stars that were obtained at the similar airmass as
the object fields are used except for
Sh 2-207 and Sh 2-208 clusters, for which 2MASS stars in the fields are
used after converting the 2MASS magnitudes to the MKO magnitudes with
the color transformations in \citet{Leggett2006}.
The achieved limiting magnitudes (10$\sigma$), $J \simeq 20$--22\,mag,
$H \simeq 19$--21\,mag, and $K_S \simeq 18.5$--21\,mag, correspond to
$\lesssim$0.1\,$M_\odot$.
This mass detection limit, sufficiently less than 1\,$M_\odot$, enables
a direct comparison of disk fractions in such distant clusters with
those of star-forming clusters in the solar neighborhood
\citep[cf.][]{Haisch2001ApJL}.
For each embedded cluster, we identified cluster members in the same way
as for the Cloud 2 clusters \citepalias{My2009ApJ}:
the extinction criterion ($A_V$) for identifying cluster members was
determined for each cluster considering the $A_V$ distributions in the
cluster region and the field region \citepalias[see details
in][]{My2009ApJ}.
The details for each cluster will be published in separated papers.

%%%%%%%%%%%%%%%%%%%%%%%%%%%%%%%%%%%%%%%%%%%%%%%%%%%%%%%%%%%%%%%%%%%%%%%%%%%%%%%

%\begin{tabular}
\begin{deluxetable*}{llllll}
\tabletypesize{\scriptsize}
\tablecaption{$JHK$ disk fractions of embedded clusters in low-metallicity
environments. \label{tab:allRESULTS}}
\tablehead{
\colhead{Cluster} & \colhead{log [O/H]$^a$} & \colhead{$R_g$ /$D^b$} & 
\colhead{Age} & \colhead{Disk Fraction} & \colhead{$M_{\rm lim}$$^c$} \\   
\colhead{} & \colhead{(dex)} & \colhead{(kpc)} & \colhead{(Myr)} &
 \colhead{(\%)} & \colhead{($M_\odot$)}
}
\startdata
 Cloud 2-N & $-0.7$ (1) & 19/12 (i) &
 0.5--1 & 9$\pm$4 $(5/52)$ & 0.06 \\ 
 Cloud 2-S & $-0.7$ (1) & 19/12 (i) &
 0.5--1 & 27$\pm$7  $(16/59)$ & 0.06 \\
 Sh 2-207 & $-0.7$ (2) & 12/4$^d$ (ii) & 2--3  & 5.1$\pm$4.6
 $(1.2/23.4)$$^e$ & 0.07--0.08 \\  
 Sh 2-208 & $-0.8$ (2) & 12/4$^d$ (ii) & 0.5 & 19$\pm$5.5
 $(12/63)$ & 0.05 \\  
 Sh 2-209 main$^f$ & $-0.6$ (2) & 17.5/10 (iii)
 & 0.5--1 & 10$\pm$0.8 $(163/1605)$ & 0.08--0.09 \\ 
 Sh 2-209 sub$^f$ & $-0.6$ (2) & 17.5/10 (iii)
 & 0.5--1 & 7.1$\pm$1.2 $(35/494)$ & 0.08--0.09
%\multicolumn{6}{c}{(\em See Online Journal Article for Full Data Table)}
\enddata
\tablecomments{
$^a$Metallicity of the observed regions.  The solar metallicity is
assumed as $\log {\rm (O/H)}_\odot = 8.7$
\citep{Asplund2009}.
$^b${Adopted distance ($D$) and Galactic distance ($R_g$). We assumed
 the Galactic distance of the sun as $R_\odot = 8.0$\,kpc \citep{Reid1993}.}
$^c${Mass detection limit of the data.}
$^d${For both Sh 2-207 and Sh 2-208 clusters, we adopted the kinematic
 distances because model KLFs with the photometric distances ($D\simeq
 8$--9\,kpc, $R_g \simeq 15.5$--16.5\,kpc) do not match the observed KLF
 at all, while those with the kinematic distances ($D\simeq 4$\,kpc)
 match very well.}
$^e${See detail in the main text.}
$^f$Sh 2-209 has two clusters, which are located near the two peaks of
millimeter continuum \citep{Klein2005}: hereafter we call the larger
cluster at the northeast side and the smaller cluster at the southwest
side as ``main cluster'' and ``sub-cluster'', respectively.
}
\tablerefs{(1) \citet{Lubowich2004}, 
(2) \citet{Caplan2000};
(i) \citet{NK2008},
(ii) \citet{Fich1990},
(iii) \citet{Chini1984}.}
%%% %
\end{deluxetable*}

\begin{figure}[!h]
\begin{center}
\includegraphics[scale=0.5]{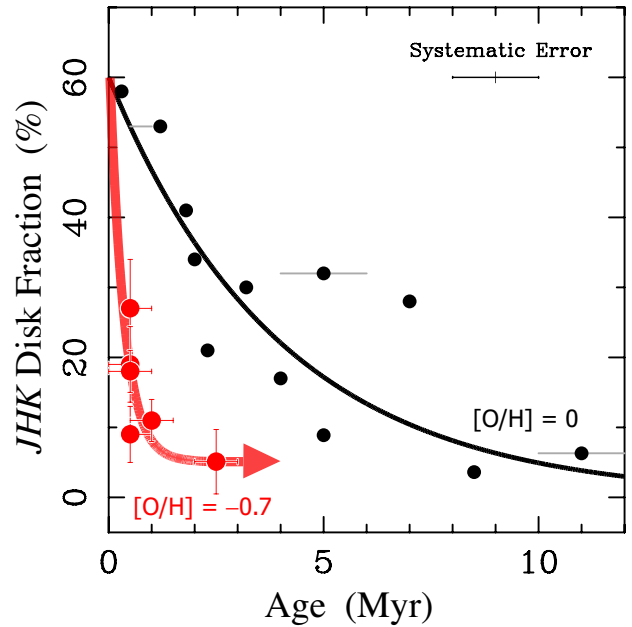}
\caption{Disk fraction as a function of cluster age.
$JHK$ disk fractions of the young clusters with low metallicity are
shown by red filled circles, while those of young clusters with solar
metallicity are shown by black filled circles.
The typical systematic error of ages for the solar metallicity cluster
is shown at the top-right corner (see the main text for explanation of
the gray horizontal lines for three of the clusters).
The black line shows the disk fraction evolution under solar
metallicity, while the red arrow shows the proposed $JHK$ disk fraction
evolution in low-metallicity environments.
Note that these lines are drawn by eye and are not derived with rigorous
fitting.}
\label{fig:DF_LM}
\end{center}
\end{figure}
%%%%%%%%%%%%%%%%%%%%%%%%%%%%%%%%%%%%%%%%%%%%%%%%%%%%%%%%%%%%%%%%%%%%%%%%%%%%%%%

\section{Derivation of Age and Disk Fraction}\label{sec:results}
\subsection{Clusters in Low Metallicity} \label{sec:Rlow}

Following the method by \citet{Muench2000}, we first derived the ages of
clusters by comparing observed the $K$-band luminosity function (KLF)
with model KLFs of various ages constructed from the canonical initial
mass function (IMF), mass--luminosity ({\it M}--{\it L}) relation and
distance of the cluster.
The detail of the model fitting is described in \citetalias{My2009ApJ}
in the case of Cloud 2. We applied the same procedure to all clusters.
We confirmed that the variation of the IMF and {\it M}--{\it L} relation
in the relevant metallicity range is small enough so that the ages of
the clusters can be estimated within the uncertainty of $\lesssim$1\,Myr
\citep{My2006ApJ}, which is sufficient for the present study.
Estimated ages of most targets are found to be very young, $\sim$1\,Myr,
but one cluster, Sh 2-207, shows a relatively old age (2--3\,Myr), as
expected from the 2MASS image, which shows a less-extincted cluster.

Next, we used $JHK$ color--color diagrams to derive the disk
fractions for each cluster.
The method we used to determine cluster infrared excess fractions from
color--color diagrams is described in detail in \citetalias{My2009ApJ}
along with extensive discussions on the uncertainties.
In Table~\ref{tab:allRESULTS}, we list the derived disk fractions and
ages for the entire sample clusters, and plotted the results in
Figure~\ref{fig:DF_LM}.
The uncertainties of the disk fraction reflect Poisson errors.
Only for the Sh 2-207 cluster we derived the disk fraction by
considering the contamination from the field stars because it is
difficult to identify cluster members with the small extinction of this
older cluster.
The number of stars in the Sh 2-207 cluster region is 38 and of these
only 2 have disks.
The normalized number of stars in the field region is estimated to be
14.6$\pm$0.9 with that lying in the infrared excess region of the
diagrams to be 0.8$\pm$0.2. 
Therefore, the final disk fraction of the Sh 2-207 cluster is estimated
at 5.1\,\% $\pm$ 4.6\,\% $((2-0.8) / (38-14.6))$.

%%%%%%%%%%%%%%%%%%%%%%%%%%%%%%%%%%%%%%%%%%%%%%%%%%%%%%%%%%%%%%%%%%%%%%%%%%%%%%%

%\begin{table*}
\begin{table}
\begin{center}
\caption{{\it JHK} disk fractions of embedded clusters with ages of
 $\ge$5\,Myr in the solar neighborhood.}\label{tab:DF}
%\begin{tabular}{cccccccc}
\begin{tabular}{lllllll}
\hline
\hline
%\tablenotemark{1}
% ID & Cluster & Age$^{\rm a}$ & $JHK$ disk fraction & Cluster & Age\footnotemark & DF &  Mass\footnotemark &
%Cluster &  Age & Disk Fraction\footnotemark &  Mass\footnotemark &
Cluster & Age$^a$ & Disk Fraction &  $M_{\rm lim}$$^b$ & Filter$^c$ &
 Ref$^d$ \\  
 & (Myr) & (\%) &  ($M_\odot$) & \\
\hline
 Orion OB1b & 5 (1) & 8.9$\pm$3 & 0.1 & 2MASS & 1\\
% Orion OB1b & 4--6 (1) & 8.9$\pm$3 & 0.1 & 2MASS & 1\\
%Briceño+ 2005 \\
%
%
 Upper Sco & 5 (2) & 32$\pm$4 & 0.1 & 2MASS & 3 \\
% Upper Scorpius & 5 (2) & 32$\pm$4 & 0.1 & 2MASS & 2 \\
%※ Carpenter2006 \\
%
%
 $\eta$ Cham & 6 (4) & 28$\pm$12 & 0.08 & 2MASS & 5 \\
% $\eta$ Cham & 5--9 (3) & 28$\pm$12 & 3.2--0.08 & 2MASS & 4 \\
%Luhman \& Steeghs (2004) \\
%
%
 Orion OB1a & 8.5 (1) & 3.6$\pm$2.5 & 0.1 & 2MASS & 1 \\
%\Briceno+ 2005 \\
%
%
 NGC 7160 & 10 (5) & 6.3$\pm$3.6 & 0.4 & 2MASS & 6 \\
%Sicilia-Aguilar+ 2006 ApJ\\
%
\hline
\end{tabular}
\end{center}
\tablecomments{
%\tablenotetext{a}
%{N{\scriptsize OTES.---}
$^{a}$References for the ages are shown in the parenthesis. 
Although there are two age estimates for Orion OB1b and OB1a by using
different isochrone models, we show the average value here.
$^b$Mass detection limit of the data.
$^c$The photometric system of the obtained $JHK$ data.
$^d$References for the $JHK$ photometric data. 
For Upper Scorpius and NGC 7160, we used $JHK$ magnitudes in 2MASS
All-Sky Point Source Catalog for the stars listed in the references.}
\tablerefs{
(1) \citet{Briceno2005},
(2) \citet{Preibisch2002},
(3) \citet{Carpenter2006},
%(4) \citet{Megeath2005},
(4) \citet{LuhmanSteeghs2004},
(5) \citet{Sicilia-Aguilar2005ApJ130},
%(6) \citet{Sicilia-Aguilar2004},
(6) \citet{Sicilia-Aguilar2006}.}
\vspace{5mm}
\end{table}
%%%%%%%%%%%%%%%%%%%%%%%%%%%%%%%%%%%%%%%%%%%%%%%%%%%%%%%%%%%%%%%%%%%%%%%%%%%%%%%

\subsection{Reference Clusters in Solar Metallicity}
For comparison, we also derived the $JHK$ disk fraction and age of
various embedded clusters in the solar neighborhood {\it in a uniform
way} using the photometry data in the literature.
We chose publications with the following criteria:
(1) $JHK$ photometry data set of cluster members in the same filter
system are available for a reliable estimate
and (2) the mass detection limit is $\le$1\,$M_\odot$
\citep{Haisch2001ApJL} to sufficiently cover the most typical mass range
in order to have the most representative disk fraction for each cluster.
In addition to seven young ($<5$\,Myr) clusters in the solar
neighborhood used in \citetalias{My2009ApJ} (see Table~\ref{tab:DF}), we
derived disk fractions for additional five clusters with older ages
($\ge$5\,Myr; Orion OB1b, Upper Scorpius, $\eta$ Cham, Orion OB1a, and
NGC 7160; see Table~\ref{tab:DF}).

Although the ages of these clusters are derived with other methods than
KLF fitting (e.g., with spectroscopy and imaging data using the H-R
diagram), we derived the ages of the clusters with KLF fitting for
better consistency.
Because the KLF fitting is reliable only for good signal-to-noise
imaging data which cover the peak magnitude of KLF, we derived ages of
Trapezium, Upper Sco, and NGC 7160 clusters, since these are the only
clusters for which we could find such 
%good signal-to-noise 
data in the literature.
The resultant ages are identical to those in the literature in spite of
larger uncertainties with increasing ages as shown with thick gray
horizontal lines in Figure~1.
Considering that ages of reference clusters estimated with various
pre-main-sequence models have systematic uncertainty of $\sim$1\,Myr
\citep{Haisch2001ApJL}, the age estimate of clusters with
low-metallicity with KLF fitting in $\S$~\ref{sec:Rlow} appears to be
reliable.

For all clusters, the $JHK$ disk fractions were estimated in the same
way as for the clusters in low-metallicity environments.
The resultant disk fractions and ages for solar-metallicity clusters are
shown in Figure~1 with black circles.
They show decreasing disk fraction with increasing age: the decline is
rapid up to 5--7\,Myr and stays almost flat beyond this point with disk
fraction slowly declining from 10\,\% to 5\,\%.
These characteristics are totally identical to the latest MIR disk
fractions based on ground-based $L$-band observations and {\it Spitzer}
observations \citep[see Figure~14 in][]{Hernandez2007}.

\section{DISK LIFETIME IN LOW-METALLICITY ENVIRONMENTS} \label{sec:DL}

Figure~1 clearly shows that the disk fractions of all the clusters in
low-metallicity environments are less than 30\,\%, which is
significantly lower than nearby embedded clusters of similar age. 
Also, the disk fraction with low metallicity seems to reach the first
lowest level ($\sim$10\,\%) at the age of $\sim$1\,Myr, which
corresponds to the disk fraction with solar metallicity at
$\sim$5--7\,Myr.
In combination with the fact that there are no embedded clusters of
solar metallicity whose disk fractions are this low at the same ages,
our results show that disk fraction strongly depends on metallicity,
confirming our earlier suggestion in \citetalias{My2009ApJ}.

Following the detailed discussion in \citetalias{My2009ApJ}, our results
suggest that the {\it entire} disk dispersal occurs in the very early
phase in low-metallicity environments, while initially (at ${\rm
age}=0$\,Myr) an inner disk as well as outer disk exists even under low
metallicity.
The disk lifetime in low metallicity is estimated at $\sim$1\,Myr from
Figure~1, while that in solar metallicity is $\sim$5--7\,Myr, suggesting
that very short disk lifetime in low-metallicity environment.
Assuming that the disk lifetime--metallicity relation is described as
$\tau_{\rm disk} \propto 10^{aZ}$, the constant $a$ is estimated to be
$1 \pm 0.5$ despite that we have only two measurement points (0\,dex and
$-0.7$\,dex) at present.
The uncertainty of the constant comes from the uncertainties of
metallicity and age of each cluster.
Note that our NIR observations are sensitive to only heated dust in the
inner disk (see detail in $\S$~1) and millimeter/sub-millimeter
observations of the outer disk (e.g., with ALMA) in the near future are
very important to conclude that the above estimated disk lifetime is
actually for the entire disk.

The metallicity dependence of the disk lifetime may impose constraints
on the mechanism of disk dispersal, which is still under debate even for
solar metallicity environments.
Among five major mechanisms of disk dispersal:
planet formation, stellar encounter, stripping by winds, mass accretion,
and photoevaporation \citep{Hollenbach2000PPIV},
only photoevaporation by the radiation of the central star appears to
cause the strong metallicity dependence (see discussion in
\citetalias{My2009ApJ}; \citealt{Ercolano2010}) if the dominant
radiation is in far-UV (FUV, $6\,{\rm eV} < h\nu < 13.6$\,eV; e.g.,
\citealt{Gorti2009a}) and/or X-rays as suggested in
\citetalias{My2009ApJ} ($h\nu > 0.1$\,keV; e.g.,
\citealt{Ercolano2010}).
However, because the theory of photoevaporation is still under
development and with numerous parameters, further studies are needed.
The metallicity dependencies of disk lifetime from theories are
$\propto$$10^{0.3Z}$ and $\propto$$10^{0.5Z}$ for FUV photoevaporation
from the theories of \citet{Gorti2009a} and X-ray photoevaporation from
those of \citet{Ercolano2010}, respectively.
This is not enough to explain our results, although it is marginally
consistent within the uncertainty ($10^{aZ}$, $a=1\pm0.5$).
Note that something specific to the outer Galaxy environment
\citep[e.g.,][]{Haywood2008} might be the reason for the rapid disk
dispersal.
However, considering that Sh 2-207 and Sh 2-208 clusters, which are not
found to be located in the outer Galaxy region ($R_g = 12$\,kpc, see
note in Table~1), also show low disk fraction and thus short disk
lifetime, metallicity could be the dominant factor.

\section{IMPLICATION TO THE PLANET--METALLICITY CORRELATION}

The observed planet--metallicity correlation is known to have strong
metallicity dependence ($P_{\rm pl} \propto 10^{2Z}$;
\citealt{Fischer2005}).
Although the core-accretion model qualitatively explains the dependence,
e.g., $P_{\rm pl} \propto 10^Z$ in the case of \citet{Ida2004ApJ616},
which is a deterministic model based on the core accretion, it is still
not enough to explain the steep $10^{2Z}$ dependence (see e.g.,
\citealt{Wyatt2007}).
The short disk lifetime under low-metallicity suggests that planet
formation becomes more difficult with decreasing metallicity, and
therefore it may explain a part of the strong planet--metallicity
correlation.
In \citetalias{My2009ApJ}, we proposed that by adding the contribution
from the disk lifetime, most of the observed metallicity dependence of
the planet formation probability could be reasonably explained.
Recently, \citet{Ercolano2010} also estimated the
correlation between $P_{\rm pl}$ and $\tau_{\rm disk}$ by combining
their photoevaporation model and the core-accretion model by
\citet{Ida2004ApJ616}
to suggest that the metallicity dependence of photoevaporation only
plays a secondary role with an effect of $\lesssim 10^{0.2Z}$ on the
planet formation probability.
However, there are more ambiguities and complexities in the planet
formation process than the theoretical planet formation scenarios now
available \citep[e.g.,][]{Ida2004ApJ604}. 
Also, the derived metallicity dependence of the disk lifetime still has
a large uncertainty (see $\S$~\ref{sec:DL}), and needs to be confirmed
by observations of clusters with higher than solar metallicity.
The metallicity dependence of the disk lifetime suggested here is the
only direct observational evidence for the metallicity dependence of the
planet formation and could be an important clue to understanding the
planet formation mechanism.

%%%%%%%%%%%%%%%%%%%%%%%%%%%%%%%%%%%%%%%%%%%%%%%%%%%%%%%%%%%%%%%%%%%%%%%%%%%%%%%

\acknowledgments

C.Y. has been supported by the Japan Society for the Promotion of
Science (JSPS).

%%%%%%%%%%%%%%%%%%%%%%%%%%%%%%%%%%%%%%%%%%%%%%%%%%%%%%%%%%%%%%%%%%%%%%%%%%%%%%%

%%%%%%%%%%%%%%%%%%%%%%%%%%%%%%%%%%%%%%%%%%%%%%%%%%%%%%%%%%%%%%%%%%%%%%%%%%%%%%%

%\clearpage

%%% !
%\tablerefs{(1) \citet{Meyer1997}, (2) \citet{Haisch2000AJ}, (3)
%\citet{Palla2000}, (4) \citet{Muench2002}, (5) \citet{Luhman2004}, (6)
%\citet{Kenyon1995}, (7) \citet{Haisch2001AJ}, (8) \citet{Rebull2002},
%(9) \citet{Sicilia-Aguilar2005ApJ130},

% !!! %
%%% (i) \citet{Muzzio1974}, (ii) \citet{Smartt1996},
%%% (iii) \citet{Digel1994}, (iv) \citet{Stil2001}, 
%%% (v) \citet{Moffat1979}, 
%%% (vi) \citet{Chini1984}, (vii) \citet{Pismis1991}, 
%%% (viii) \citet{Fich1990},
%%% (ix) \citet{Caplan2000},
%%% (x) \citet{Lahulla1985}.}

\end{document}